\documentclass[5p,times]{elsarticle}
\usepackage{lineno,hyperref}
\modulolinenumbers[100]

\journal{Journal of \LaTeX\ Templates}
\usepackage{graphicx}
\graphicspath{ {images/} }
\usepackage{subfig}
\usepackage{float}
\usepackage{amsmath}
\usepackage{epstopdf}
\begin{document}

\begin{frontmatter}
\title{Phase stability and elastic properties in the Al\textsubscript{1-x-y}Cr\textsubscript{x}Ti\textsubscript{y}N   system from first principles}
\tnotetext[mytitlenote]{Pie de imagen}

\author{Erik Guti\'{e}rrez-Valladares $^{1}$, Rurick Santos-Fragoso  $^{1,2}$, Guillermo V\'{a}zquez-Tovar $^{1,2}$, Andr\'{e}s Manuel Garay-Tapia  $^{3}$, Diego Germ\'{a}n Espinosa-Arbel\'{a}ez $^{1}$, Raymundo Arr\'oyave $^{4}$, Jes\'{u}s Gonz\'{a}lez-Hern\'{a}ndez $^{1}$, Juan Manuel Alvarado-Orozco  $^{1}$}
\address{
$^{1}$ \quad Direcci\'{o}n de Ingenier\'{i}a de Superficies, Centro de Ingenier\'{i}a y Desarrollo Industrial, Quer\'{e}taro, 76125, M\'{e}xico.; egutierrez@posgrado.cidesi.edu.mx\\
$^{2}$ \quad Facultad de Ingenier\'{i}a, Universidad Aut\'{o}noma de Quer\'{e}taro, Quer\'{e}taro, 76230, M\'{e}xico\\
$^{3}$ \quad Centro de Investigaci\'{o}n en Materiales Avanzados, Unidad Monterrey, 66600, Apodaca, Nuevo Le\'on, M\'{e}xico.\\
$^{4}$ \quad Department of Materials Science and Engineering, Mechanical Engineering,  Texas A$\&$M University College of Engineering, College Station, Texas, USA.\\}


\begin{abstract}
Multicomponent nitrides are a hot research topic in the search of hard coatings. The effect of substitutions on the phase stabilities, magnetic, and elastic properties of Al\textsubscript{1-x-y}Cr\textsubscript{x}Ti\textsubscript{y}N (x,y=0-1) was studied using first principles calculations based on the density functional theory.  These calculations provide the lattice parameter, formation energy, mixing enthalpy and elastic constants. The calculated values  are in good agreement with experiments and compare well with other theoretical results. A magnetic transition from antiferromagnetic to ferromagnetic state occurs  at concentrations of  B1-TiN higher than 60\%. The quaternary zone has a lower aluminum solubility than the constituent ternary systems. The Poisson's ratio, Shear and Young modulus were used as predictors of the hardness, indicating that the higher hardness values are found on the transition line from B1 to B4. The obtained results  enable the design of new Al\textsubscript{1-x-y}Cr\textsubscript{x}Ti\textsubscript{y}N-based materials for coating  applications.   
\end{abstract}


\end{frontmatter}

\linenumbers

\section{Introduction}

Transition metal nitrides have been widely used as protective coatings for machining, forging, and stamping applications due to their outstanding physical properties such as high hardness, chemical inertness, thermal stability, as well as oxidation and wear resistance \cite{Veprek1999a,Hultman2000}. 
TiN and CrN coatings were an improvement compared to the traditional methods of heat treatment and nitriding,enchancing hardness to 28 and 20 GPa,  and operating temperatures up to 550 and 700 $^{\circ}$C,  respectively \cite{Chim2009b}. 

CrN presents an antiferromagnetic behavior below the N\'{e}el temperature ($\sim$280 K)  with an alternated spin  double-layer configuration on an orthorhombic structure with $\alpha$=88.23$^{\circ}$ \cite{Alling2007b}.
Above this temperature, CrN becomes paramagnetic with B1 structure \cite{Corliss1960}.
The role of the magnetic contribution in Cr-containing systems is still an open research topic \cite{Steneteg2012a,Alling2010,Alling2013a,Alling2007b,Mayrhofer2008c}. 

In order to improve the hardness, oxidation resistance, cutting performance, wear resistance, and the operating temperature more complex multicomponent nitrides have been studied. 
Particularly, it was discovered that by adding Al to TiN and CrN  the above mentioned characteristics are improved  \cite{Willmann2008a, Endrino2006b,Kumar2014b,Reiter2005a, Yamamoto,Schaffer2000}. 
AlCrN and AlTiN are reported to be stable in the NaCl (B1) structure under an Al \% at. composition --usually reported to be around 70\%--, whereas at larger Al content occurs a transition to hexagonal (B4) structure which has a detrimental effect on the hardness of the coating \cite{Reiter2005a,Alling2011}.
Comparing both ternary nitrides, AlCrN exhibits a slightly lower hardness than AlTiN at the same Al content \cite{Alling2007b}.
However, Cr-based ternary nitrides can dissolve a greater amount of Al in the B1 structure which retains the  stability of the system to a higher temperature (925$^{\circ}$C) \cite{Willmann2006}.
In spite of Al having a strong influence on high temperature stability of the ternary nitrides, the Ti- and Cr- content grants the systems  different degradation mechanisms.
For AlCrN nucleation and growth is the main responsible for high temperature degradation ($\sim$1000$^{\circ}$C); whereas for AlTiN the degradation mechanism ($\sim$900$^{\circ}$C) is related to spinodal decomposition which involves the transformation of metastable cubic AlTiN zones into TiN-rich and AlN-rich zones  \cite{Alling2007b,Mayrhofer2005b,Alling2007f}.




Currently, there are plenty experimental and theoretical studies of AlCrN and AlTiN in which structural stability, mechanical properties, elastic constants, and magnetic behavior have been studied. 
However, only a few experimental and theoretical studies of CrTiN were found \cite{Alling2010,Inumaru2007}.
CrTiN has lower hardness values across its different compositions compared to all other ternary nitride systems \cite{Hones1998}. In spite of that drawback, the surprising magnetic behavior and the capacity of Cr and Ti to form protective oxide layers (Cr\textsubscript{2}O\textsubscript{3} and TiO\textsubscript{2}) are interest for electrical  and corrosion protective applications \cite{Alling2010,Choi2009,Inumaru2007,Inumaru2008,Inumaru2004}.

Even though the previously discussed ternary systems are still an open research topic, the quaternary system Al\textsubscript{1-x-y}Cr\textsubscript{x}Ti\textsubscript{y}N has received attention due to the improvement of the cutting performance, hardness, and wear resistance at high temperature compared to the ternary nitrides. \cite{Polcar2011,Xu2015,Endrino2005a,Forsen2013d}. 
Experimental studies have found that the oxidation resistance is negatively affected by Ti concentrations higher than 11 at. \% due to the promotion of a TiO$_{2}$ porous surface layer over the more protective Al$_{2}$O$_{3}$ \cite{Forsen2013d}.
The pseudo-ternary AlCrTiN presents also a spinodal decomposition similar to AlTiN, where  Cr is distributed into two metastable zones (AlCrN and TiCrN) resulting in an age hardening process \cite{Forsen2012}. Furthermore, Cr addition in the AlN phase decelerates the formation of the wurtzite phase, enhancing the thermal stability compared to the pseudo-binary TiAlN or CrAlN alloys \cite{Forsen2012,Hugosson2003,Lind2011}. 
 






Several experimental research focused on specific concentration points of this quaternary nitride can be found but only few theoretical studies focus on a full concentration range with emphasis on spinodal decomposition \cite{Lind2011,Xu2017,Zhou2017}. 


In this contribution, the Density Functional Theory (DFT) and the Special Quasirandom Structure (SQS) technique  were employed to model Al\textsubscript{1-x-y}Cr\textsubscript{1-x}Ti\textsubscript{1-y}N alloys. 
what makes this multicomponent nitride  interesting is its chemically complexity and special rules must be followed for the atoms distribution into the lattice sites in order to represent the magnetic state. 
The study presents a complete calculation of a pseudo-ternary system using concentration steps of 12.5\%. The configurations proposed in this study to simulate the magnetic effects are just an approximation of the most stable state. 
The stress-strain method was used to calculate the stiffness matrix on relaxed SQSs to accurately represent the alloy local effects \cite{Shin2006a,Shin2007}. 




The aim of this paper is to study the quaternary nitride AlCrTiN obtaining the Al solubility limit, elastic constants trends, phase stability, and second derivative of the mixing energy to observe the probability of spinodal decomposition.

\section{Computational Methodology}



The SQSs were made using the Alloy Theoretic Automated Toolkit (ATAT) software, which is based on the method proposed by Zunger \textit{et al.}, for representing a valid model of Al\textsubscript{1-x-y} Cr\textsubscript{1-x}Ti\textsubscript{1-y}N in a small supercell computationally feasible for DFT calculations \cite{Atat2013,VandeWalle2009, AlexZungerS.-H.WeiL.G.Ferreira1990}.
The theoretical calculations are performed within the projector augmented wave (PAW) method and the density functional theory framework (DFT) as implemented in the Vienna ab-initio simulation package (VASP) \cite{Blochl1994, Kohn1965,Jones1989, Kresse1993,Kresse1996, Kresse1996a}. The electronic exchange-correlation potential is described within the generalized-gradient approximation (GGA) parametrized by Perdew \textit{et al} \cite{Perdew1996a}. 
The integration over the Brillouin zone employed the Monkhorst-Pack scheme using a grid of $5 \times 5 \times 5$ k-points for the B1 structures and a $9 \times 9 \times 4$ for the B4 structures \cite{Pack1977}.  An  cut-off energy of 600eV was used for the calculations. The cells for both B1 and B4 structures went through a series of volume and ionic relaxations.

The Al solubility limit for B4 structures is expected to be above 50.0\% Al. Therefore, for the B4 crystalline structures only concentrations above 50.0\% Al were calculated, except for the AlCrN and AlTiN ternary nitrides in which all concentrations points were studied.
 
\begin{figure}[H]
\includegraphics[width=8cm]{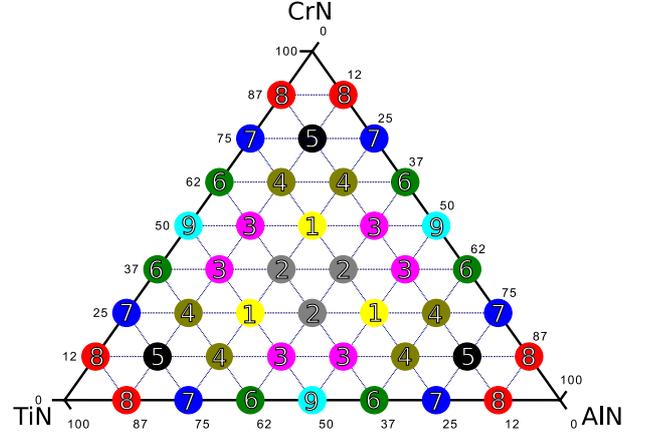}
\centering
\caption{Phase Diagram of AlCrTiN}
\label{PD}
\centering
\end{figure}

Figure \ref{PD} shows the calculated compositions represented with circles. The grid consists of 45 concentration points with composition steps of 12.5 \%, each point has a number label indicating which of the 9 generated SQSs was used.


The non-magnetic, ferromagnetic, and antiferromagnetic states at each concentration for the cubic and the hexagonal phases were calculated to find the most stable magnetic arrangement.  For the the Cr-containing points, the system was simulated as  $Al_{1-x-y}Cr_{x/2}\uparrow Cr_{x/2}\downarrow Ti_{y}N $. Although an orthorhombic antiferromagnetic calculation was performed in the case of CrN, the reported states is the antiferromagnetic with Cr atoms arranged in an alternating spin single-layer configuration.  


When the SQSs were relaxed, the total energy $E$ (obtained from VASP) was compared against the most stable phases of each alloy component to obtain the formation enthalpy. For a quaternary nitride, the formation enthalpy $\Delta H_f$ is defined as
\begin{equation}
\begin{matrix}
n\Delta H_f = E \left ( Al_{1-x-y}Cr_xTi_y N \right ) \\ - \left ( 1 -x - y\right )nE \left (Al\right ) -xnE \left(Cr \right) -ynE \left(Ti \right) - \left( n/2 \right) E \left(N_2 \right)
\end{matrix}
\label{for}
\end{equation}
In equation \ref{for} the total energy of N was calculated for a N$_2$ molecule, while the total energies of Al, Cr and Ti were calculated for their unit cells in their respective stable structure.

Mixing Enthalpy $\Delta H_{m}$
of cubic Al\textsubscript{1-x-y}Cr\textsubscript{x}Ti\textsubscript{y}N is defined as

\begin{equation}
\begin{matrix}
n\Delta H_m = E \left( Al_{1-x-y} Cr_x Ti_y N \right) \\- \left( 1-x-y \right) \left( n/2 \right) E \left( AlN \right) -x \left(n/2 \right) E \left( CrN \right) -y \left(n/2 \right) E \left( TiN \right)
\end{matrix}
\label{mix}
\end{equation}

 In equation \ref{mix} the energies $E$ of AlN, CrN and TiN are calculated in cubic phase.



\subsection{Elastic Constants}

The elastic constants were calculated using the stress-strain method 
\cite{LePage2002,Shang2007}. Positive and negative strains with a magnitude of $x=\pm 0.01$ were applied in all the strain tensor directions to modify the relaxed SQSs. 
Subsequently, the pseudo-inverse method to obtain the elastic constants values was applied  and the stiffness tensor was calculated for the stable magnetic state of each SQS \cite{Shang2007}. 

Since the point group symmetry is broken by the SQS approach, the elastic tensor matrix is symmetric and  contains 21 values.
The 9 non-zero values 
were averaged to obtain the macroscopic cubic elastic constants: 
$\overline{C_{11}}=\left ( 1/3 \right )\left (   C_{11}+C_{22}+C_{33}\right )$,  $\overline{C_{12}}=\left ( 1/3 \right )\left (   C_{12}+C_{13}+C_{23}\right )$, $\overline{C_{44}}=\left ( 1/3 \right )\left (   C_{44}+C_{55}+C_{66}\right )$ 
\cite{Moakher2006}.





\section{Results}

\subsection{Magnetic Stability}
\begin{figure}[H]
\includegraphics[width=8.5cm]{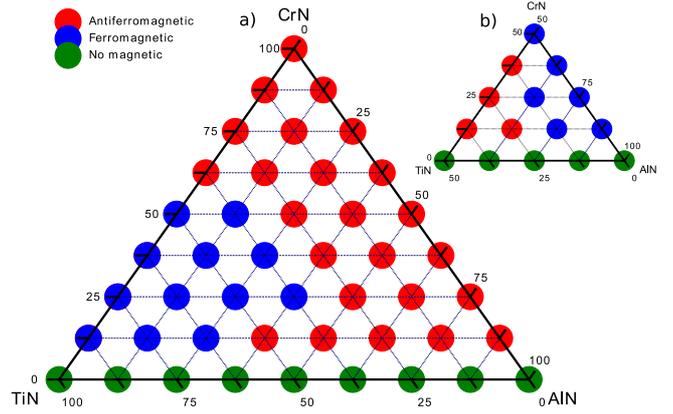}
\centering
\caption{Magnetic Stability}\label{MS}
\centering
\end{figure}

The magnetic properties of Al\textsubscript{1-x-y}Cr\textsubscript{x}Ti\textsubscript{y}N in the complete range of composition for the cubic phase and the hexagonal phase for Al content $>50$\% are shown in Figure \ref{MS}a and \ref{MS}b, respectively. 

The TiN, AlN, and pseudo binary TiAlN do not exhibit magnetic behavior in any of the compositions of the cubic and hexagonal crystalline phases calculated \cite{Alling2007b}. As mentioned above, CrN presents an AFM behavior (below the Neel temperature) relating this with a distortion from cubic phase to the orthorhombic phase with a double layer arrangement [110]. In this study the AFM [100] arrangement is used  to maintain the cubic phase \cite{Corliss1960}. 

The pseudo binary B1-AlCrN presents an AFM behavior in the complete range of compositions and B1-TiCrN exhibits a change in the magnetic behavior from FM to AFM.  In this study due to the approximation used (GGA), the point B1-Ti\textsubscript{50}Cr\textsubscript{50}N presents an energy difference between the AFM and FM arrangements of 1 meV/atom making it difficult to determine the composition in which the transition takes place. According to Alling \textit{et al.}, who use the LDA+U approximation, this change in the magnetic behavior occurs when there is approximately 
50 at. content\cite{Alling2010}. Because of this,  the point B1-Ti\textsubscript{50}Cr\textsubscript{50}N is presented in the figure 4a  as FM  based on the observed tendency of the energy difference of the points B1-Ti\textsubscript{62.5}Cr\textsubscript{37.5}N and B1-Ti\textsubscript{37.5}Cr\textsubscript{62.5}N, which agrees with the magnetic transition reported by Alling \textit{et al.} 

For B1-Al\textsubscript{1-x-y}Cr\textsubscript{x}Ti\textsubscript{y}N a shift can also be observed  in the magnetic behavior from AFM to FM. In the quaternary points studied, only the points where  TiN$\leq$25 \% at. content as well as Al\textsubscript{50}Cr\textsubscript{12.5}Ti\textsubscript{37.5}N, and Al\textsubscript{37.5}Cr\textsubscript{12.5}Ti\textsubscript{50}N present an AFM behavior. It is important to mention that the energy difference between the AF and AFM states of the  Al\textsubscript{12.5}Cr \textsubscript{50}Ti\textsubscript{37.5}N and Al\textsubscript{25}Cr\textsubscript{12.5}Ti\textsubscript{62.5}N concentration points is similar to the difference in the B1-Ti\textsubscript{50}Cr\textsubscript{50}N  point  (1 meV/atom) making it difficult to determine the most stable magnetic state for these points,  and the observed tendencies of the points are taken into account to determine the behavior represented in Figure \ref{MS}a.  

The points with CrN =12.5\% at. content may be influenced by the position of the two Cr atoms in the SQSs since, as reported by Filippetti \textit{et al.} \cite{Filippetti1998,Filippetti2000}, in the CrN the interactions (J) of the first neighbors have a negative value, favoring the AFM state while the interactions of second neighbors have a positive value favoring the FM state. This low CrN content can also explain the reason of the minimal energy difference between the FM and AFM states for the Al\textsubscript{25}Cr\textsubscript{12.5}Ti\textsubscript{62.5}N. 

Regarding the hexagonal phase, the pseudo binary B4-AlCrN presents a magnetic change from FM to AFM when there is CrN$\geq $63 \% at. content (see Figure \ref{MS}b) which is in agreement with previous reports \cite{Zhang2005,Wu2003}. Likewise, for the quaternary B4-Al\textsubscript{1-x-y}Cr\textsubscript{x} Ti\textsubscript{y}N, the same magnetic shift was observed as in B4-AlCrN. The three quaternary points with Al=50\% at. content and the point B4-Al\textsubscript{62.5}Cr\textsubscript{12.5}Ti\textsubscript{25}N present an AFM behavior while the remaining points were FM.

\subsection{Lattice Parameter}
The lattice parameters were calculated from the relaxed SQSs shown in Figure \ref{PD}; based on them, color map graphs were obtained by interpolation.  

\begin{figure}[H]
\includegraphics[width=8.5cm]{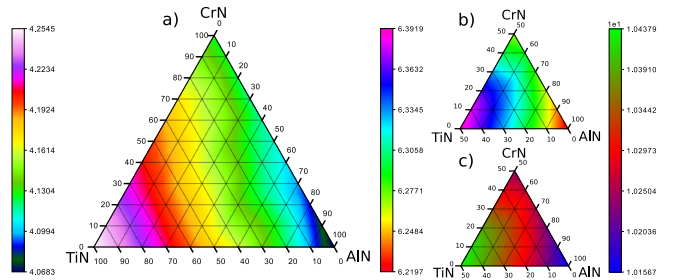}
\centering
\caption{Lattice Parameter\label{Lat Par}}
\centering
\end{figure}

The Figure \ref{Lat Par} shows, the lattice parameter variation for the cubic and hexagonal phases. The figure 2a is the ternary diagram for the cubic phase with its scale located at the left.
The triangles in the right correspond to the lattice parameters of the hexagonal phase, figure \ref{Lat Par}b correspond to the parameter \textit{a} with the scale located at the center and figure \ref{Lat Par}c to the parameter \textit{c} with its scale located at the right. 

The lattice parameter of the cubic phase increases almost as if it was only dependent on the Ti content. This may be promoted by the relatively big lattice parameter of TiN in comparison with the smaller lattice parameters of AlN and CrN. The same tendency in lattice parameter was found for the hexagonal phase where the maximum value for \textit{a} and \textit{c} was found at the TiN, while the minimum was located at the AlN point.



Even the lattice parameter change in the cubic phase for Al content higher than 50\% has the same greater dependency on Cr content as the lattice parameters in the hexagonal phase have. 



\subsection{Formation Enthalpy}

\begin{figure}[H]
\includegraphics[width=8.5cm]{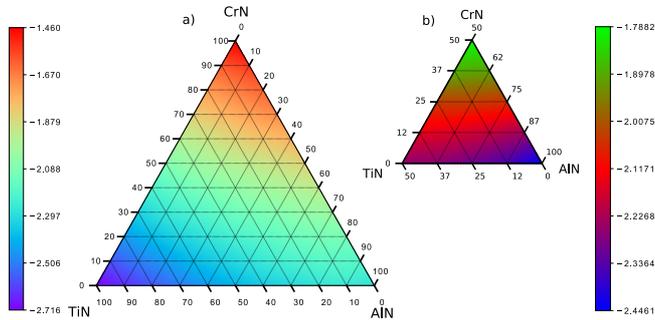}
\centering
\caption{Formation Enthalpy\label{FE}}
\centering
\end{figure}

The Formation Enthalpy for the cubic and hexagonal phase are presented in Figure \ref{FE}a and \ref{FE}b respectively, the equation \ref{for} shows how the values for Formation Enthalpy were obtained. 

The AFM B1-CrN presents the highest formation enthalpy value of the entire ternary diagram including the hexagonal phase (Figure \ref{FE}b), the formation enthalpy obtained (-1.460 eV/at) is in good agreement with the value reported by Mayrhofer \textit{et al.} [\cite{Mayrhofer2008c}]. The B1-TiN presents the lowest formation enthalpy value of the entire ternary diagram and the AlN an intermediate value between the B1-CrN and B1-TiN.

For the pseudo binary nitrides B1-AlTiN and B1-CrAlN the obtained results are also in agreement with the values reported by Alling \textit{et al.} and Mayhorfer \textit{et al.}, for the case of B1-TiCrN no reported values were found, but it follows the trend of the other two pseudo binaries.

The formation enthalpy values of the quaternary nitride B1-AlTiCrN do not show a tendency established by any of the secondary nitrides, rather we can notice how the ternary diagram is divided into 3 zones, the corner of the TiN, the corner of the CrN and a zone that goes from the corner of the AlN to the line of the TiCrN nitride, the formation enthalpies of the B1-AlCrTiN are mainly in the middle of the scale values. 

For B4-AlN as it was expected we obtain a lower formation enthalpy than B1-AlN, also it can be appreciated a change in the enthalpy minimum from the B1-TiN corner to the B4-AlN corner. The trends for the B4-AlCrTiN are not inclined to any of the secondary nitride, but with more B4-AlN content the B4 phase is more likely to be the most stable phase. And, if we maintain the Al and Ti content relation, more B4-CrN content means a higher formation energy.  This behavior has found also in the B1 phase, only that B4 has a steepest ascent. Only comparing values from both formation energies we can calculate the Al solubility limit. 

\subsection{Phase stability}

\begin{figure}[H]
\includegraphics[width=8.5cm]{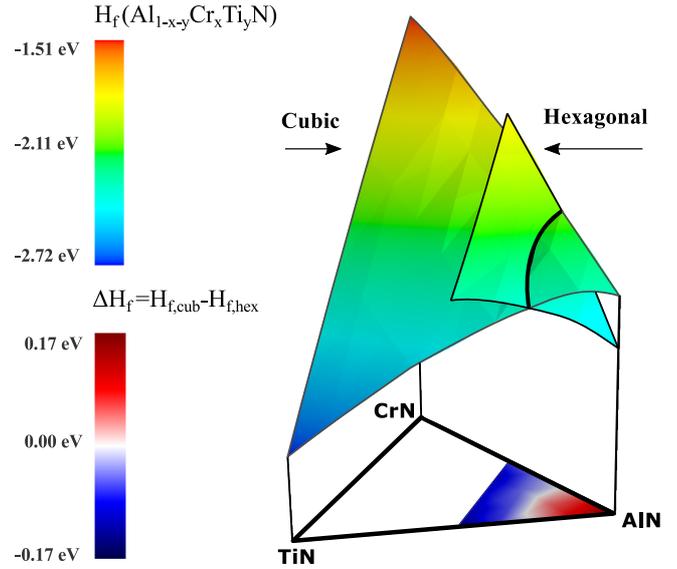}
\centering
\caption{Phase Stability\label{PhaseStab}}
\centering
\end{figure}

The formation energy of the cubic and hexagonal structures were plotted as 3D ternary surfaces in Figure \ref{PhaseStab}, the intersection of both surfaces show where the cubic to the hexagonal transformation occurs. The projection shows the difference of formation energy between the cubic and the hexagonal phase $\Delta H_f=H_{f,cub}-H_{f,hex}$. 
The blue color in the Figure \ref{PhaseStab} shows the stable zone of cubic phase and the transition line point out the aluminium solubility limit.  
The aluminum solubility in the quaternary compositions is lower than the ternary nitrides (AlCrN, AlTiN).  

\subsection{Mixing Enthalpy}
\begin{figure}[H]
\includegraphics[width=8.5cm]{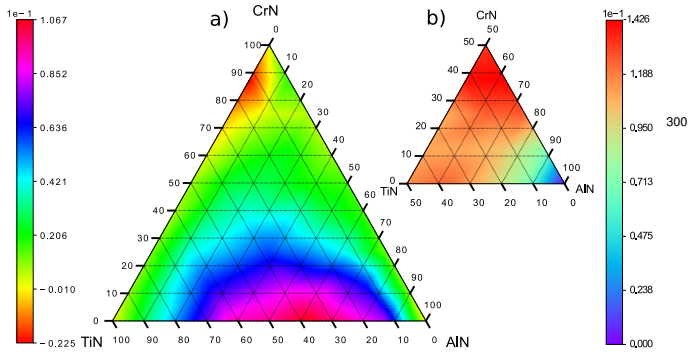}
\centering
\caption{Mixing Enthalpy}\label{EMez}
\centering
\end{figure}

The Mixing Enthalpy for the cubic and hexagonal phase are presented in Figure \ref{EMez}a and \ref{EMez}b respectively. This values were obtained using the equation \ref{mix} explained in the section above.  
The ternary nitrides are in good agreement with the values previously reported by Alling et.al. and Mayhorfer et.al. \cite{Mayrhofer2006a,Alling2007b,Alling2007f}  
In Figure \ref{EMez}a it can be seen the same tendency found in the ternary nitrides, also a negative mixing enthalpy values can be found in the quaternary B1-AlCrTiN region near CrN suggesting a good stability. The B1-AlCrTiN section near the line of the B1-TiAlN is where  the highest values of Mixing Enthalpy can be found, making this the most unstable region of the quaternary.

In general, B1-CrN content lowers the value of the Mixing Enthalpy. Considering only this chemical enthalpy, it means higher Cr content nitrides could have a greater thermal stability.

While B1-CrN content seems to considerably affect the Mixing Enthalpy, it only does if B1-AlN and B1-TiN content both are present in considerable amount. If not true, the enthalpy value does not vary in a noticeable way, maintaining an almost constant region in most of the B1-AlCrTiN mixing enthalpy, this being in agreement with the values reported by Lind et. al. \cite{Lind2011} 

For B4-AlCrTiN it can be notice three evident regions shown in the figure \ref{EMez}b were the highest values can be found when there is more B4-CrN content. Only when there is more than 80\% B4-AlN we can see an improvement in the B4 phase stability.

\subsection{Elastic Constants}

\begin{figure}[H]
\includegraphics[width=8.5cm]{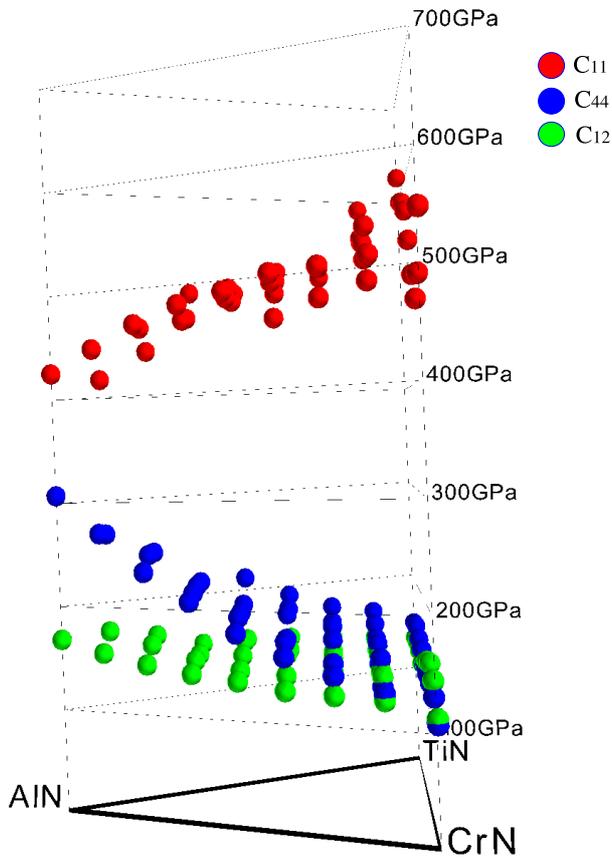}
\centering
\caption{Elastic Constants\label{ElasCons}}
\centering
\end{figure}

The elastic constants shown in fig. \ref{ElasCons} show the three macroscopic elastic constants for the Al\textsubscript{1-x-y}Cr\textsubscript{x}Ti\textsubscript{y}N alloy, $C_{11}$ (red), $C_{12}$ (green)  and $C_{44}$(blue). 
The $C_{11}$ constant has a maximum at the binary TiN, from this point the values of the ternary nitrides AlTiN and CrTiN have a negative slope reaching a minimum in points of low Ti concentration. This tendency is also present in the quaternary nitride. That is to say, low Ti concentration points show the smallest values in all the concentration region. Fixing the Ti content, the elastic constant magnitude increase with the Cr content. This is also true for AlCrN where a steady increase is present from AlN to CrN.

$C_{44}$ is not affected by the variations of Cr and Ti content, by keeping the Al content fixed the slope of the resulting lines is close to zero. While increasing the aluminum content the elastic value grows rapidly and reaches its peak at the AlN point. $C_{12}$ has the more straightforward trend of the 3 constants as it has practically the same value for all concentration points. 
The difference in the values obtained for the $C_{12}$ constant are relatively small when are compared to the other elastic constants values. However, these results show similar tendencies to the ones presented in the $C_{44}$ constant where the Al content is the most important variable for the elastic constant increment.

The Born stability criteria for the cubic system are the following conditions, which are necessary and sufficient to assert elastic stability \cite{Mouhat2014}:
\begin{equation}
\begin{matrix}
C_{11} - C_{12}>0, &  C_{11}+2C_{12}>0, & C_{44}>0
\end{matrix}
\end{equation}

This criteria were calculated for all concentrations  and were fulfilled by each one of them, showing that the structures are mechanically stable.

From the elastic tensor defined above, a number of  derived properties are calculated \cite{Hill1952,Liu2014,Fulcher2012}.





Shear modulus Voigt average, \(G_{V}\):

\[ G_{V} = 15 \left( C_{11}-C_{12}+C_{44}\right) \]


Shear modulus Reuss average, \(G_{R}\)

\[ G_{R}=\frac{5C_{44} \left ( C_{11} - C_{12} \right )}{C_{44}} \]




Shear modulus  average, \(G\)

\[2 G = \left(G_{V} + G_{R} \right)\]


We may further obtain the  the isotropic Young's modulus
\[ E=\frac{9KG}{3K+G} \]

and the isotropic Poisson's ratio

\[\nu =\frac{3K-2G}{2\left ( 3K+G \right )}\]

where  $K = \frac{1}{3} (C_{11} + 2C_{12}) $ and is the isotropic bulk modulus.


\begin{figure}[H]
\includegraphics[width=7cm]{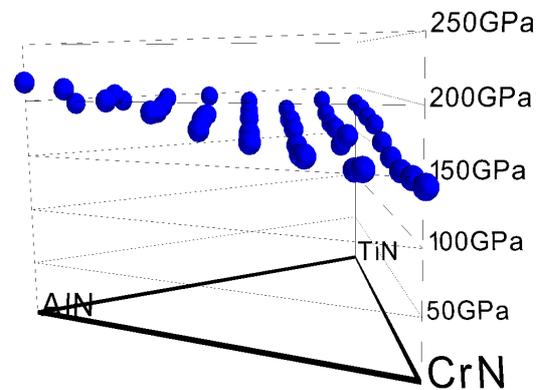}
\centering
\caption{Shear Modulus Average Trends\label{Shear}}
\centering
\end{figure}

Shear, Bulk  and Young modulus have been used as empiric predictors of hardness in general \cite{Zhou2012, Pokluda2015a}. For cubic nitrides shear modulus, the inverse of Poisson's ratio and the ideal strength were found to have the biggest correlation values \cite{Fulcher2012}. Considering this we analyze the tendencies of this constants in order to find a possible hardness tendency of the AlCrTiN system.

\begin{figure}[H]
\includegraphics[width=7cm]{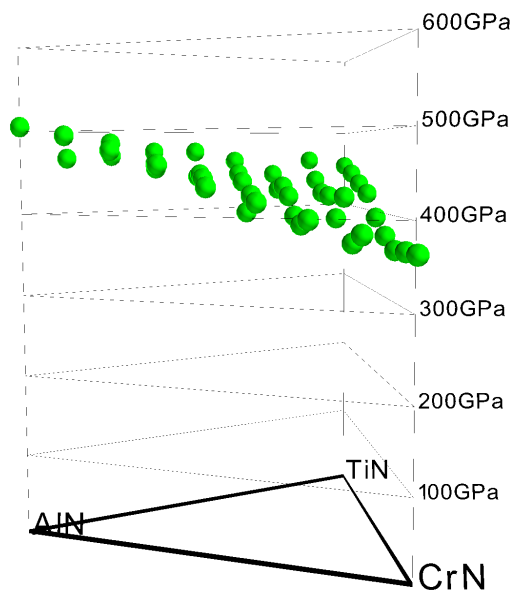}
\centering
\caption{Young Modulus Trends\label{Young}}
\centering
\end{figure}

The shear modulus biggest influencing factor is $C_{44}$, and accordingly to what can be observed in figure \ref{ElasCons} and \ref{Shear} both grow as aluminum concentration increases, so being as close to as possible to AlN would give us the higher hardness values, however, we have to take into consideration the phase stability of the cubic phase as a limiting factor, showing that the higher hardness values would be found along the transition line between the B1 and B4 phases. The existence of this line with similar hardness values gives room to play with the quantities of Ti and Cr to also consider the phase stability and oxide resistance.

\begin{figure}[H]
\includegraphics[width=6cm]{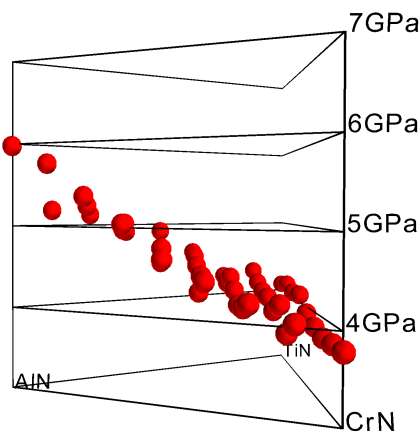}
\centering
\caption{Poisson's Ratio Trends\label{Poisson}}
\centering
\end{figure}

\section{Conclusions}
A study of the Al\textsubscript{1-x-y}Cr\textsubscript{x}Ti\textsubscript{y}N  system in the pseudo-ternary concentration diagram ($0\leq x\leq 1$ and $0 \leq y \leq 1$, with 12.5\% x and y increments) was carried out. We found a good agreement between our calculations and the results of previously published studies for the formation energy, magnetic stability and elastic properties of the binary and ternary nitrides. These results allow us to ensure that the calculations of the quaternary region are a good approximation of the above-described characteristics.

The ternary nitrides present three different magnetic states (AFM, FM and NM),  in the quaternary zone (AlCrTiN) two of this behaviors (AFM and FM) can be found as well as a change of behavior from AFM to FM when there is more than 60\% of B1-TiN content. The results presented are a good approximation of the behavior of the system, but to improve the accuracy, calculations could be made with other approximations (LDA+U). For the B4-AlCrTiN the behavior present in figure \ref{MS}b is in good agreement with the experimental results presented in previous  studies \cite{Zhang2005, Wu2003}. 

The stress-strain method was successfully used for the elastic constants calculus, all of these values are in the same range and followed characteristic trends in each constant. In this system, the approximated macroscopic elastic constants are of special importance as an aid to experimental design of materials. Since hardness can not be calculated directly from DFT calculations, we use the  Poisson's ratio, Shear and Young Modulus to predict the hardness tendencies, which indicate that the higher hardness values for cubic are expected to be found along the transition line from B1 to B4 structure. According to the difference of formation energies there will be with a bit more aluminum solubility on the AlTiN side of the transition line, and the mixing energy shows a higher stability on the AlCrN side.

\section{Acknowledgments}
The authors are grateful with CONACYT for the financial support granted through the scholarship programs \textit{Laboratorio Nacional de Proyecci\'on T\'ermica (293429)} and  \textit{Fronteras de la Ciencia (2015-02-1077)}, which made the development of the scientific activities reported possible.



\begin{thebibliography}{60}
\expandafter\ifx\csname url\endcsname\relax
  \def\url#1{\texttt{#1}}\fi
\expandafter\ifx\csname urlprefix\endcsname\relax\def\urlprefix{URL }\fi
\expandafter\ifx\csname href\endcsname\relax
  \def\href#1#2{#2} \def\path#1{#1}\fi

\bibitem{Veprek1999a}
S.~Vep{\v{r}}ek, {{The
  search for novel, superhard materials}}, Journal of Vacuum Science {\&}
  Technology A: Vacuum, Surfaces, and Films 17~(5) (1999) 2401--2420.
\newblock
  {\path{doi:10.1116/1.581977}}.

\bibitem{Hultman2000}
L.~Hultman, {Thermal stability of nitride thin films}, Vacuum 57~(1) (2000)
  1--30.
\newblock
  {\path{doi:10.1016/S0042-207X(00)00143-3}}.

\bibitem{Chim2009b}
Y.~Chim, X.~Ding, X.~Zeng, S.~Zhang,
  {{Oxidation
  resistance of TiN, CrN, TiAlN and CrAlN coatings deposited by lateral
  rotating cathode arc}}, Thin Solid Films 517~(17) (2009) 4845--4849.
\newblock
  {\path{doi:10.1016/j.tsf.2009.03.038}}.

\bibitem{Alling2007b}
B.~Alling, T.~Marten, I.~a. Abrikosov, a.~Karimi, {Comparison of thermodynamic
  properties of cubic Cr1-x Alx N and Ti1-x Alx N from first-principles
  calculations}, Journal of Applied Physics 102~(4).
\newblock
  {\path{doi:10.1063/1.2773625}}.

\bibitem{Corliss1960}
L.~M. Corliss, N.~Elliott, J.~M. Hastings, {Antiferromagnetic structure of
  CrN}, Physical Review 117~(4) (1960) 929--935.
\newblock
  {\path{doi:10.1103/PhysRev.117.929}}.

\bibitem{Steneteg2012a}
P.~Steneteg, B.~Alling, I.~a. Abrikosov, {Equation of state of paramagnetic CrN
  from ab initio molecular dynamics}, Physical Review B 85 (2012) 1--7.
\newblock
  {\path{doi:10.1103/PhysRevB.85.144404}}.

\bibitem{Alling2010}
B.~Alling, {Theory of the ferromagnetism in Ti1-x CrxN solid solutions},
  Physical Review B - Condensed Matter and Materials Physics 82~(5).
\newblock
  {\path{doi:10.1103/PhysRevB.82.054408}}.

\bibitem{Alling2013a}
B.~Alling, L.~Hultberg, L.~Hultman, I.~A. Abrikosov, {Strong electron
  correlations stabilize paramagnetic cubic Cr 1-xAlxN solid solutions},
  Applied Physics Letters 102~(3).
\newblock
  {\path{doi:10.1063/1.4788747}}.

\bibitem{Mayrhofer2008c}
P.~H. Mayrhofer, D.~Music, T.~Reeswinkel, H.~G. Fu{\ss}, J.~M. Schneider,
  {Structure, elastic properties and phase stability of Cr1-xAlxN}, Acta
  Materialia 56~(11) (2008) 2469--2475.
\newblock
  {\path{doi:10.1016/j.actamat.2008.01.054}}.

\bibitem{Willmann2008a}
H.~Willmann, P.~H. Mayrhofer, L.~Hultman, C.~Mitterer, {Hardness evolution of
  Al–Cr–N coatings under thermal load}, Journal of Materials Research
  23~(11) (2008) 2880--2885.
\newblock
  {\path{doi:10.1557/JMR.2008.0366}}.

\bibitem{Endrino2006b}
J.~L. Endrino, G.~S. Fox-Rabinovich, C.~Gey, {Hard AlTiN, AlCrN PVD coatings
  for machining of austenitic stainless steel}, Surface and Coatings Technology
  200~(24) (2006) 6840--6845.
\newblock 
  {\path{doi:10.1016/j.surfcoat.2005.10.030}}.

\bibitem{Kumar2014b}
T.~S. Kumar, S.~B. Prabu, G.~Manivasagam, K.~a. Padmanabhan, {Comparison of
  TiAlN , AlCrN , and AlCrN / TiAlN coatings for cutting-tool applications},
  International Journal of Minerals, Metallurgy, and Materials 21~(8) (2014)
  796--805.
\newblock 
  {\path{doi:10.1007/s12613-014-0973-y}}.

\bibitem{Reiter2005a}
A.~E. Reiter, V.~H. Derflinger, B.~Hanselmann, T.~Bachmann, B.~Sartory,
  {Investigation of the properties of Al1-xCrxN coatings prepared by cathodic
  arc evaporation}, Surface and Coatings Technology 200~(7) (2005) 2114--2122.
\newblock 
  {\path{doi:10.1016/j.surfcoat.2005.01.043}}.

\bibitem{Yamamoto}
K.~Yamamoto,
  {{Properties
  of (Ti,Cr,Al)N coatings with high Al content deposited by new plasma enhanced
  arc-cathode}}, Surface and Coatings Technology 174-175 (2003) 620--626.
\newblock 
  {\path{doi:10.1016/S0257-8972(03)00580-2}}.

\bibitem{Schaffer2000}
E.~Schaffer, G.~Kleer, {Mechanical behavior of (Ti , Al) / N coatings exposed
  to elevated temperatures and an oxidative environment}, Surface {\&} Coatings
  Technology 133-134 (2000) 215--219.

\bibitem{Alling2011}
B.~Alling, a.~V. Ruban, a.~Karimi, L.~Hultman, I.~a. Abrikosov, {Unified
  cluster expansion method applied to the configurational thermodynamics of
  cubic Ti1-xAlxN}, Physical Review B - Condensed Matter and Materials Physics
  83~(10) (2011) 1--21.
\newblock
  {\path{doi:10.1103/PhysRevB.83.104203}}.

\bibitem{Willmann2006}
H.~Willmann, P.~Mayrhofer, P.~Persson, A.~Reiter, L.~Hultman, C.~Mitterer,
  {{Thermal
  stability of Al–Cr–N hard coatings}}, Scripta Materialia 54~(11) (2006)
  1847--1851.
\newblock 
  {\path{doi:10.1016/j.scriptamat.2006.02.023}}.

\bibitem{Mayrhofer2005b}
P.~H. Mayrhofer, C.~Mitterer, H.~Clemens, {Self-organized nanostructures in
  hard ceramic coatings}, Advanced Engineering Materials 7~(12) (2005)
  1071--1082.
\newblock
  {\path{doi:10.1002/adem.200500154}}.

\bibitem{Alling2007f}
B.~Alling, a.~V. Ruban, A.~Karimi, O.~E. Peil, S.~I. Simak, L.~Hultman, I.~a.
  Abrikosov, {Mixing and decomposition thermodynamics of c- Ti1-x Alx N from
  first-principles calculations}, Physical Review B - Condensed Matter and
  Materials Physics 75~(4) (2007) 1--13.
\newblock
  {\path{doi:10.1103/PhysRevB.75.045123}}.

\bibitem{Inumaru2007}
K.~Inumaru, K.~Koyama, Y.~Miyaki, K.~Tanaka, S.~Yamanaka,
  {{Ferromagnetic
  Cr(x)Ti(1-x)N solid solution nitride thin films grown by pulsed laser
  deposition and their magnetoresistance}}, Applied Physics Letters 91~(15)
  (2007) 152501.
\newblock 
  {\path{doi:10.1063/1.2776853}}.

\bibitem{Hones1998}
P.~Hones, R.~Sanjin{\'{e}}s, F.~L{\'{e}}vy, {Sputter deposited chromium nitride
  based ternary compounds for hard coatings}, Thin Solid Films 332~(1-2) (1998)
  240--246.
\newblock 
  {\path{doi:10.1016/S0040-6090(98)00992-4}}.

\bibitem{Choi2009}
H.~S. Choi, D.~H. Han, W.~H. Hong, J.~J. Lee, {(Titanium, chromium) nitride
  coatings for bipolar plate of polymer electrolyte membrane fuel cell},
  Journal of Power Sources 189~(2) (2009) 966--971.
\newblock 
  {\path{doi:10.1016/j.jpowsour.2008.12.060}}.

\bibitem{Inumaru2008}
K.~Inumaru, Y.~Miyaki, K.~Tanaka, K.~Koyama, S.~Yamanaka,
  {{Magnetoresistance of
  ferromagnetic CrTiN solid solution nitride}}, Physical Review B 78~(5) (2008)
  052406.
\newblock 
  {\path{doi:10.1103/PhysRevB.78.052406}}.

\bibitem{Inumaru2004}
K.~Inumaru, T.~Ohara, K.~Tanaka, S.~Yamanaka, {Layer-by-layer deposition of
  epitaxial TiN-CrN multilayers on MgO(0 0 1) by pulsed laser ablation},
  Applied Surface Science 235~(4) (2004) 460--464.
\newblock 
  {\path{doi:10.1016/j.apsusc.2004.03.260}}.

\bibitem{Polcar2011}
T.~Polcar, A.~Cavaleiro,
  {{Structure and
  tribological properties of AlCrTiN coatings at elevated temperature}},
  Surface and Coatings Technology 205~(SUPPL. 2) (2011) S107--S110.
\newblock
  {\path{doi:10.1016/j.surfcoat.2011.03.015}}.

\bibitem{Xu2015}
Y.~Xu, L.~Chen, Z.~Liu, F.~Pei, Y.~Du,
  {{Influence
  of Ti on the mechanical properties, thermal stability and oxidation
  resistance of Al–Cr–N coatings}}, Vacuum 120 (2015) 127--131.
\newblock
  {\path{doi:10.1016/j.vacuum.2015.07.004}}.

\bibitem{Endrino2005a}
J.~L. Endrino, V.~Derflinger, {The influence of alloying elements on the phase
  stability and mechanical properties of AlCrN coatings}, Surface and Coatings
  Technology 200~(1-4 SPEC. ISS.) (2005) 988--992.
\newblock 
  {\path{doi:10.1016/j.surfcoat.2005.02.196}}.

\bibitem{Forsen2013d}
R.~Fors{\'{e}}n, M.~P. Johansson, M.~Od{\'{e}}n, N.~Ghafoor,
  {{Effects of Ti alloying of
  AlCrN coatings on thermal stability and oxidation resistance}}, Thin Solid
  Films 534~(534) (2013) 394--402.
\newblock 
  {\path{doi:10.1016/j.tsf.2013.03.003}}.

\bibitem{Forsen2012}
R.~Fors{\'{e}}n, M.~Johansson, M.~Od{\'{e}}n, N.~Ghafoor,
  {{Decomposition and
  phase transformation in TiCrAlN thin coatings}}, Journal of Vacuum Science
  {\&} Technology A: Vacuum, Surfaces, and Films 30~(6) (2012) 061506.
\newblock 
  {\path{doi:10.1116/1.4757953}}.

\bibitem{Hugosson2003}
H.~W. Hugosson, H.~H{\"{o}}gberg, M.~Algren, M.~Rodmar, T.~I. Selinder, {Theory
  of the effects of substitutions on the phase stabilities of Ti1-xAlxN},
  Journal of Applied Physics 93~(8) (2003) 4505--4511.
\newblock 
  {\path{doi:10.1063/1.1557779}}.

\bibitem{Lind2011}
H.~Lind, R.~Fors{\'{e}}n, B.~Alling, N.~Ghafoor, F.~Tasndi, M.~P. Johansson,
  I.~A. Abrikosov, M.~Od{\'{e}}n, {Improving thermal stability of hard coating
  films via a concept of multicomponent alloying}, Applied Physics Letters
  99~(9).
\newblock 
  {\path{doi:10.1063/1.3631672}}.

\bibitem{Xu2017}
Y.~X. Xu, H.~Riedl, D.~Holec, L.~Chen, Y.~Du, P.~H. Mayrhofer, {Thermal
  stability and oxidation resistance of sputtered Ti[sbnd]Al[sbnd]Cr[sbnd]N
  hard coatings}, Surface and Coatings Technology 324 (2017) 48--56.
\newblock 
  {\path{doi:10.1016/j.surfcoat.2017.05.053}}.

\bibitem{Zhou2017}
J.~Zhou, L.~Zhang, L.~Chen,
  {{Effect of Cr on
  metastable phase equilibria and spinodal decomposition in c-TiAlN coatings: A
  CALPHAD and Cahn-Hilliard study}}, Surface and Coatings Technology 311 (2017)
  231--237.
\newblock 
  {\path{doi:10.1016/j.surfcoat.2017.01.007}}.

\bibitem{Shin2006a}
D.~Shin, R.~Arr{\'{o}}yave, Z.~K. Liu, A.~Van De~Walle, {Thermodynamic
  properties of binary hcp solution phases from special quasirandom
  structures}, Physical Review B - Condensed Matter and Materials Physics
  74~(2).
\newblock
  {\path{doi:10.1103/PhysRevB.74.024204}}.

\bibitem{Shin2007}
D.~Shin, A.~Van De~Walle, Y.~Wang, Z.~K. Liu, {First-principles study of
  ternary fcc solution phases from special quasirandom structures}, Physical
  Review B - Condensed Matter and Materials Physics 76~(14) (2007) 1--10.
\newblock 
  {\path{doi:10.1103/PhysRevB.76.144204}}.

\bibitem{Atat2013}
A.~van~de Walle, M.~Asta, G.~Ceder,
  {{The
  alloy theoretic automated toolkit: A user guide}}, Calphad 26~(4) (2002)
  539--553.
\newblock 
  {\path{doi:10.1016/S0364-5916(02)80006-2}}.

\bibitem{VandeWalle2009}
A.~van~de Walle, {Multicomponent multisublattice alloys, nonconfigurational
  entropy and other additions to the Alloy Theoretic Automated Toolkit},
  Calphad: Computer Coupling of Phase Diagrams and Thermochemistry 33~(2)
  (2009) 266--278.
\newblock 
  {\path{doi:10.1016/j.calphad.2008.12.005}}.

\bibitem{AlexZungerS.-H.WeiL.G.Ferreira1990}
A.~Zunger, S.-H. Wei, L.~G. Ferreira, J.~E. Bernard,
  {{Special
  quasirandom structures}}, Physical Review Letters 65~(3) (1990) 353--356.
\newblock 
  {\path{doi:10.1103/PhysRevLett.65.353}}.

\bibitem{Blochl1994}
P.~E. Bl{\"{o}}chl, {Projector augmented-wave method}, Physical Review B
  50~(24) (1994) 17953--17979.
\newblock 
  {\path{doi:10.1103/PhysRevB.50.17953}}.

\bibitem{Kohn1965}
W.~Kohn, L.~J. Sham, {Self-consistent equations including exchange and
  correlation effects}, Physical Review 140~(4A).
\newblock 
  {\path{doi:10.1103/PhysRev.140.A1133}}.

\bibitem{Jones1989}
R.~O. Jones, O.~Gunnarsson, {The density functional formalism, its applications
  and prospects}, Reviews of Modern Physics 61~(3) (1989) 689--746.
\newblock 
  {\path{doi:10.1103/RevModPhys.61.689}}.

\bibitem{Kresse1993}
G.~Kresse, J.~Hafner, {Ab initio molecular dynamics for liquid metals},
  Physical Review B 47~(1) (1993) 558--561.
\newblock 
  {\path{doi:10.1103/PhysRevB.47.558}}.

\bibitem{Kresse1996}
G.~Kresse, J.~Furthm{\"{u}}ller, {Efficiency of ab-initio total energy
  calculations for metals and semiconductors using a plane-wave basis set},
  Computational Materials Science 6~(1) (1996) 15--50.
\newblock 
  {\path{doi:10.1016/0927-0256(96)00008-0}}.

\bibitem{Kresse1996a}
G.~Kresse, J.~Furthm{\"{u}}ller, {Efficient iterative schemes for ab initio
  total-energy calculations using a plane-wave basis set}, Physical Review B -
  Condensed Matter and Materials Physics 54~(16) (1996) 11169--11186.
\newblock 
  {\path{doi:10.1103/PhysRevB.54.11169}}.

\bibitem{Perdew1996a}
J.~D. Perdew, K.~Burke, M.~Ernzerhof, {Generalized Gradient Approximation Made
  Simple}, Physical Review Letters 77~(3) (1996) 3865--3868.

\bibitem{Pack1977}
J.~D. Pack, H.~J. Monkhorst, {"special points for Brillouin-zone
  integrations"-a reply}, Physical Review B 16~(4) (1977) 1748--1749.
\newblock 
  {\path{doi:10.1103/PhysRevB.16.1748}}.

\bibitem{LePage2002}
Y.~Le~Page, P.~Saxe,
  {{Symmetry-general
  least-squares extraction of elastic data for strained materials from ab
  initio calculations of stress}}, Physical Review B 65~(10) (2002) 104104.
\newblock 
  {\path{doi:10.1103/PhysRevB.65.104104}}.

\bibitem{Shang2007}
S.~Shang, Y.~Wang, Z.~K. Liu, {First-principles elastic constants of
  {$\alpha$}- And {$\theta$}- Al2O3}, Applied Physics Letters 90~(10) (2007)
  3--6.
\newblock 
  {\path{doi:10.1063/1.2711762}}.

\bibitem{Moakher2006}
M.~Moakher, A.~N. Norris, {The closest elastic tensor of arbitrary symmetry to
  an elasticity tensor of lower symmetry}, Journal of Elasticity 85~(3) (2006)
  215--263.
\newblock 
  {\path{doi:10.1007/s10659-006-9082-0}}.

\bibitem{Filippetti1998}
A.~Filippetti, W.~E. Pickett, B.~M. Klein,
 {{Competition
  between Magnetic and Structural Transition in CrN}} (1998) 1--10
  {\path{doi:10.1103/PhysRevB.59.7043}}.

\bibitem{Filippetti2000}
A.~Filippetti, N.~A. Hill, {Magnetic stress as a driving force of structural
  distortions: The case of CrN}, Physical Review Letters 85~(24) (2000)
  5166--5169.
\newblock 
  {\path{doi:10.1103/PhysRevLett.85.5166}}.

\bibitem{Zhang2005}
J.~Zhang, X.~Z. Li, B.~Xu, D.~J. Sellmyer, {Influence of nitrogen growth
  pressure on the ferromagnetic properties of Cr-doped AlN thin films}, Applied
  Physics Letters 86~(21) (2005) 1--3.
\newblock 
  {\path{doi:10.1063/1.1940131}}.

\bibitem{Wu2003}
S.~Y. Wu, H.~X. Liu, L.~Gu, R.~K. Singh, L.~Budd, M.~van Schilfgaarde, M.~R.
  McCartney, D.~J. Smith, N.~Newman,{{Synthesis, characterization,
  and modeling of high quality ferromagnetic Cr-doped AlN thin films}}, Applied
  Physics Letters 82~(18) (2003) 3047--3049.
\newblock 
  {\path{doi:10.1063/1.1570521}}.

\bibitem{Mayrhofer2006a}
P.~H. Mayrhofer, D.~Music, J.~M. Schneider, {Ab initio calculated binodal and
  spinodal of cubic Ti 1-xAl xN}, Applied Physics Letters 88~(7) (2006) 16--19.
\newblock 
  {\path{doi:10.1063/1.2177630}}.

\bibitem{Mouhat2014}
F.~Mouhat, F.~Coudert, {Necessary and Sufficient Elastic Stability Conditions
  in Various Crystal Systems}, Physical Review B 90 (2014) 224104.
\newblock 
  {\path{doi:10.1103/PhysRevB.90.224104}}.

\bibitem{Hill1952}
R.~Hill,
  {{The
  Elastic Behaviour of a Crystalline Aggregate}}, Proceedings of the Physical
  Society. Section A 65~(5) (1952) 349--354.
\newblock 
  {\path{doi:10.1088/0370-1298/65/5/307}}.

\bibitem{Liu2014}
Z.~T.~Y. Liu, X.~Zhou, S.~V. Khare, D.~Gall,
  {{Structural, mechanical
  and electronic properties of 3d transition metal nitrides in cubic
  zincblende, rocksalt and cesium chloride structures: a first-principles
  investigation.}}, Journal of physics. Condensed matter : an Institute of
  Physics journal 26~(2) (2014) 025404.
\newblock 
  {\path{doi:10.1088/0953-8984/26/2/025404}}.

\bibitem{Fulcher2012}
B.~D. Fulcher, X.~Y. Cui, B.~Delley, C.~Stampfl, {Hardness analysis of cubic
  metal mononitrides from first principles}, Physical Review B - Condensed
  Matter and Materials Physics 85~(18) (2012) 1--9.
\newblock
  {\path{doi:10.1103/PhysRevB.85.184106}}.

\bibitem{Zhou2012}
L.~Zhou, D.~Holec, P.~H. Mayrhofer,{{First-principles study of
  elastic properties of cubic Cr 1− x Al x N alloys}}, Journal of Applied
  Physics 113~(4) (2013) 043511.
\newblock
  {\path{doi:10.1063/1.4789378}}.

\bibitem{Pokluda2015a}
J.~Pokluda, M.~{\v{C}}ern{\'{y}}, M.~{\v{S}}ob, Y.~Umeno,
  {{Ab
  initio calculations of mechanical properties: Methods and applications}},
  Progress in Materials Science 73 (2015) 127--158.
\newblock
  {\path{doi:10.1016/j.pmatsci.2015.04.001}}.

\end{thebibliography}
\end{document}